\documentclass[%
 reprint,
superscriptaddress,
 amsmath,amssymb,
 aps,
 pre,
]{revtex4-2}

\usepackage{graphicx,color}
\usepackage{dcolumn}
\usepackage{bm}
\usepackage{graphicx}
\usepackage{dcolumn}
\usepackage{bm}
\usepackage{color}

\usepackage{subfigure}

\usepackage{mathptmx}  
\usepackage{setspace}  

\usepackage{orcidlink}

\usepackage{soul}
\usepackage{tikz}

\usepackage{hyperref}
\hypersetup{colorlinks=true,linkcolor=blue,urlcolor=blue,citecolor=blue}
\usepackage[toc,page]{appendix}
\usepackage[normalem]{ulem}

\usepackage{tcolorbox}
\tcbset{colframe=blue,colback=white,boxrule=1pt,arc=3pt}

\hypersetup{colorlinks = true,linkcolor = blue,anchorcolor = blue,citecolor = blue, filecolor = blue,urlcolor = blue}
\usetikzlibrary{shapes,arrows,positioning,fit,backgrounds,calc}
\usetikzlibrary{shapes.geometric, arrows}
\usepackage{hyperref}
\usepackage[mathlines]{lineno}

\begin{document}

\preprint{APS/123-QED}

\title{Equation of Motion for Thermodynamic Systems and Their Dynamical Equivalence with Mechanical and Electrical Systems}

\author{Ekrem Aydiner}
\email{ekrem.aydiner@istanbul.edu.tr}
 \affiliation{Department of Physics, İstanbul University, İstanbul 34134, Türkiye}

\date{August 26, 2026, Ankara}

\begin{abstract}

Changes in the internal energy of a thermodynamic system are fundamentally described by the first law of thermodynamics. However, the first law has a highly compact form that does not explicitly reveal the dynamics of the system or provide a corresponding equation of motion. While equations of motion can be written explicitly for mechanical and electrical systems, an analogous dynamical formulation for thermodynamic systems has, to our knowledge, not been established directly from the first law. In this work, we show that, through dimensional reduction, the first law can be transformed from its compact thermodynamic form into an explicit equation of motion for an open thermodynamic system. Remarkably, the resulting equation has the same mathematical structure as
the equations governing a driven damped harmonic oscillator and a driven
series RLC circuit. Detailed dimensional analyses of the derived equation
and the effective coefficients further confirm the dimensional consistency
of the formulation. We further perform a systematic comparison of the mechanical, electrical, and thermodynamic systems and show that, despite their distinct physical origins, they share the same universal dynamical structure. Finally, we construct a Lagrangian formulation of the thermodynamic system and establish its correspondence with the Lagrangian descriptions of the mechanical and electrical systems. These results demonstrate that an explicit equation of motion can be generated directly from the first law of thermodynamics and reveal a common dynamical framework underlying thermodynamic, mechanical, and electrical systems.

\end{abstract}

\maketitle


\section{Introduction} \label{Intro}

\begin{figure}[h!]
\centering
\includegraphics[width=0.7\linewidth]{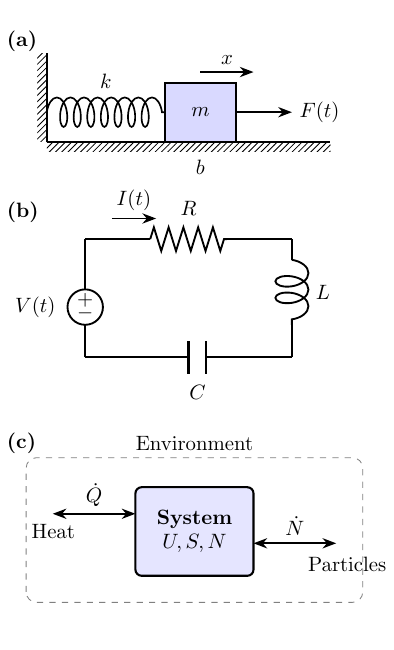}
\caption{Schematic representation of three dynamically analogous systems:
(a) a driven damped mechanical oscillator, (b) a driven series RLC circuit,
and (c) an open thermodynamic system exchanging heat and particles with its
environment.}
\label{fig:x-figure}
\end{figure}

Fig~\ref{fig:x-figure} illustrates three different physical systems:
(a) a mechanical system, (b) an electrical system, and (c) an open
thermodynamic system.

The dynamical equivalence between mechanical and electrical systems is
well known. In particular, the equations governing a classical damped
mechanical oscillator and a driven RLC circuit possess the same mathematical
structure and therefore belong to the same dynamical class. By contrast, an
analogous formulation for thermodynamic systems has not, to our knowledge,
been developed in this form. In thermodynamic systems, energy transfer and
evolution are naturally characterized through changes in the internal energy.

For the mechanical system shown in Fig.~\ref{fig:x-figure}(a), the equation
of motion is
\begin{equation}
m\ddot{x}
+
b\dot{x}
+
kx
=
F(t).
\label{clas_eq_mo}
\end{equation}
Here, $m$ is the mass, $b$ is the damping coefficient, $k$ is the spring
constant, and $F(t)$ is the external driving force.

Similarly, the dynamics of the driven RLC circuit shown in
Fig.~\ref{fig:x-figure}(b) is governed by
\begin{equation}
L\ddot{q}
+
R\dot{q}
+
\frac{1}{C}q
=
V(t),
\label{el_eq_mo}
\end{equation}
where $L$, $R$, and $C$ denote the inductance, resistance, and capacitance,
respectively, $q$ is the electric charge, and $V(t)$ is the externally
applied voltage.

Eqs~(\ref{clas_eq_mo}) and (\ref{el_eq_mo}) are structurally
equivalent: both are second-order linear dynamical equations consisting of
inertial, dissipative, restoring, and driving terms. This well-established
correspondence naturally raises an important question:
\begin{quote}
\emph{Can an analogous equation of motion be formulated for a
thermodynamic system?}
\end{quote}

In this work, we address this question. In the following sections, we
develop step by step an equation of motion for the open thermodynamic system
illustrated in Fig.~\ref{fig:x-figure}(c), which is allowed to exchange heat
and particles with its surroundings. We show that its dynamics can be
represented by an equation having the same mathematical structure as those
governing the mechanical and electrical systems shown in
Figs.~\ref{fig:x-figure}(a) and \ref{fig:x-figure}(b). Thus, despite their
distinct physical origins, the three systems can be described within a
common dynamical framework.

In Sec.~\ref{Generalized}, the contributions arising from each term of the first law are analyzed separately, and the procedure by which the first law is transformed into the required dimensional form is discussed in detail. The resulting dynamical correspondences among the mechanical, electrical, and thermodynamic systems are then summarized and presented in a comparative table.  
In Sec.~\ref{Lagrange}, the Euler--Lagrange formulation of the
thermodynamic system is developed. The kinetic- and potential-energy
terms, dissipative contributions, and power terms of the mechanical,
electrical, and thermodynamic systems are then presented and compared
in a common table. In addition, the canonical correspondence among
the three systems, including their generalized coordinates, generalized
velocities, Lagrangians, canonical momenta, and canonical pairs, is
summarized in a separate table. 
In Sec.~\ref{Con}, discussions on the main results and conclusions of the present work
are briefly summarized. The detailed dimensional analyses are provided
in Appendix, where a systematic comparison of the
mechanical, electrical, and thermodynamic systems is also presented.

\section{Derivation of the Thermodynamic Equation of Motion from the First Law}
\label{Generalized}

We begin with the fundamental thermodynamic relation for an open system,
\begin{equation}
dU
=
T\,dS
-
p\,dV
+
\mu\,dN.
\label{eq:fundamental_first_law}
\end{equation}
which known as first law of the thermodynamics. %
Here, the internal energy is regarded as a thermodynamic state function,
\begin{equation}
U=U(S,V,N),
\end{equation}
with
\begin{equation}
T=
\left(
\frac{\partial U}{\partial S}
\right)_{V,N},
\qquad
p=
-
\left(
\frac{\partial U}{\partial V}
\right)_{S,N},
\qquad
\mu=
\left(
\frac{\partial U}{\partial N}
\right)_{S,V}.
\end{equation}
Evaluating the total differential along the thermodynamic trajectory
gives
\begin{equation}
\dot U
=
T\dot S
-
p\dot V
+
\mu\dot N.
\label{eq:fundamental_power_balance}
\end{equation}
Equation~(\ref{eq:fundamental_power_balance}) follows directly from the
total differential of the state function $U(S,V,N)$, rather than from
differentiating the products $TS$, $pV$, or $\mu N$. Consequently,
terms such as $\dot T S$, $\dot p V$, and $\dot\mu N$ do not appear.

The three terms on the right-hand side represent, respectively, the
entropy, pressure--volume, and particle-transfer contributions to the
rate of change of the internal energy,
\begin{equation}
\dot {U}_{S}=T\dot {S},
\qquad
\dot {U}_{V}=-p\dot {V},
\qquad
\dot {U}_{N}=\mu\dot {N}.
\label{eq:three_thermodynamic_channels}
\end{equation}
At this stage, these expressions represent the full thermodynamic
contributions. Their dynamical parts associated with deviations from
equilibrium will be identified separately below.

In the following analysis, the three thermodynamic contributions are
treated within the same near-equilibrium linear-response framework.
The distinction between the roles of temperature, pressure, and chemical
potential reflects the separation between reference-state quantities
and dynamical deviations from equilibrium. In particular, the
temperature is evaluated at its reference equilibrium value,
$T\simeq T_0$, whereas the pressure deviation
$\delta p=p-p_0$ and the chemical-potential difference
$\Delta\mu=\mu_1-\mu_2$ are retained as dynamical quantities. The
relevant response coefficients are likewise evaluated at the reference
state and treated as constant to leading order. For notational
simplicity, the reference temperature $T_0$ will hereafter be denoted
simply by $T$.

To express the three thermodynamic contributions within a common
dynamical framework, we introduce a generalized thermodynamic function
$\phi(t)$. The coordinate $\phi$ is not identified with any particular
thermodynamic variable such as $S$, $V$, or $N$. Rather, it parametrizes
the effective thermodynamic mode considered near equilibrium, while
$\dot{\phi}$ represents the corresponding generalized rate.

The precise relation between $\phi$ and the individual thermodynamic
variables depends on the thermodynamic channel under consideration and
will therefore be specified separately in the corresponding
derivations. In this way, the same generalized coordinate $\phi$ can
be used to construct a common reduced dynamical description without
identifying it with any single thermodynamic state variable.

The three thermodynamic contributions will now be considered separately.

\subsection{Entropy Contribution: $\dot{U}_{S} ^{dyn} = T\dot{S}$ }

In extended irreversible thermodynamics (EIT) \cite{Jou_1988,Jou_2010,Mueller_Ruggeri,Israel_Stewart}, the entropy is promoted from
a function of the conventional local-equilibrium variables to a generalized
state function that may also depend explicitly on dissipative fluxes.
Schematically, one writes
$S=S\!\left(U,V,N,J \right)$
where $J$ denotes the relevant transport flux, such as a heat,
particle, or viscous flux. This extension is motivated by the fact that,
away from strict local equilibrium, the fluxes need not relax
instantaneously to their constitutive values and may therefore carry
independent dynamical information.

Near equilibrium, the entropy can be expressed in terms of the fluxes.
Because the equilibrium state corresponds to vanishing dissipative flux,
$J=0$, and the entropy must be maximal there, the term linear in
$J$ vanishes. Retaining the leading nonvanishing correction
therefore gives a quadratic contribution of the form
\begin{equation}
S
=
S_{\rm eq}
-
\frac{1}{2}\,\alpha_{J} J^{2}
+
\mathcal{O}\!\left(J^{3}\right),
\end{equation}
where $\alpha_J>0$ is a phenomenological coefficient whose precise form
depends on the transport channel and on the thermodynamic state. The
negative sign guarantees that the entropy is reduced when the system is
driven away from equilibrium by a finite dissipative flux. This quadratic
flux dependence is one of the characteristic ingredients of extended
irreversible thermodynamics and provides the basis for introducing finite
relaxation times and hyperbolic transport equations
\cite{Jou_1988,Jou_2010,Mueller_Ruggeri,Israel_Stewart}.
\begin{equation}
S = S_{\rm eq} - \frac{\alpha}{2}J^{2},
\qquad
\alpha>0,
\label{eq:EIT_entropy_expansion}
\end{equation}
where $J$ denotes the generalized flux. Eq.~(\ref{eq:EIT_entropy_expansion}) can be rewritten as
\begin{equation}
\Delta S
=
S_{\rm eq}-S
=
\frac{\alpha}{2}J^2.
\label{eq:entropy_deficit_flux}
\end{equation}
Flow of flux is maintained when
\begin{equation}
\Delta S\geq 0.
\end{equation}
We identify the generalized flux with the rate of the transfer
coordinate,
\begin{equation}
J = \dot{\phi}.
\label{eq:J_qdot}
\end{equation}
Equation~(\ref{eq:entropy_deficit_flux}) therefore becomes
\begin{equation}
\Delta S
=
\frac{\alpha}{2}\dot{\phi}^{\,2}.
\label{eq:entropy_deficit_qdot}
\end{equation}
For constant $\alpha$, differentiation with respect to time yields
\begin{align}
\dot{\Delta S} = \frac{d}{dt} \left( \frac{\alpha}{2}\do{\phi}^{2} \right) =
\alpha\dot{\phi}  \ddot{\phi}.
\label{eq:entropy_deficit_rate}
\end{align}
Thus,
\begin{equation}
\dot {S}_{\rm eq}-\dot S
=
\alpha\dot{\phi} \,\ddot {\phi}.
\label{eq:entropy_difference_rate}
\end{equation}
Multiplying by the temperature gives
\begin{equation}
T
\left(
\dot {S}_{\rm eq} - \dot {S}
\right) = \alpha T \dot{\phi} \,\ddot{\phi}.
\label{eq:entropy_power_intermediate}
\end{equation}
We now define
\begin{equation}
M \equiv\alpha T.
\label{eq:M_alphaT}
\end{equation}
The entropy-deficit contribution consequently takes the form
\begin{equation}
T\frac{d}{dt}
\left( S_{\rm eq} - S \right) = M \dot{\phi} \,\ddot{\phi}.
\label{eq:M_power_final}
\end{equation}
Therefore, the entropy-related dynamical energy-transfer rate can be
written as
\begin{equation}
\dot {U}_{S}^{\rm dyn} = T \dot{S} = M \dot{\phi} \,\ddot{\phi} .
\label{eq:entropy_dynamic_power}
\end{equation}

The physical interpretation of this contribution becomes particularly
transparent when it is compared with the corresponding inertial
energy-storage mechanisms in mechanical and electrical systems. In the
present thermodynamic description, the quadratic contribution associated
with the generalized flux $\dot{\phi}$ takes the form
\begin{equation}
U_{\rm th}^{\rm dyn} = \frac{1}{2}M\dot{\phi}^{\,2},
\end{equation}
whose time derivative is
\begin{equation}
\dot {U}_{\rm th}^{\rm dyn} = M \dot{\phi}\ddot{\phi}.
\end{equation}
This term represents energy stored in the dynamical transport mode itself.
Within the framework of extended irreversible thermodynamics, such a
contribution originates from treating the transport flux as an independent
dynamical variable with a finite relaxation time. The corresponding
coefficient $M$ therefore characterizes an effective thermodynamic inertia
associated with changes of the generalized flux.

The same mathematical structure appears in a mechanical system, where the
kinetic energy is
\begin{equation}
U_{\rm kin}
=
\frac{1}{2}m\dot{x}^{\,2},
\qquad
\dot {U}_{\rm kin}
=
m\dot{x}\ddot{x},
\end{equation}
and in an electrical circuit, where the magnetic energy stored in an
inductor is
\begin{equation}
U_{L} = \frac{1}{2}LI^2.
\end{equation}
Introducing the electric charge $Q$ through $I=\dot Q$, this can be written
as
\begin{equation}
U_{L} = \frac{1}{2}L\dot Q^{\,2},
\qquad
\dot {U}_{L} = L\dot {Q}\,\ddot {Q}.
\end{equation}
Hence, the three systems exhibit the same quadratic inertial
energy-storage structure,
\begin{equation}
\frac{1}{2}M\dot{\phi}^{\,2} \quad\longleftrightarrow\quad \frac{1}{2}m\dot{x}^{\,2}
\quad\longleftrightarrow\quad
\frac{1}{2}L\dot {Q}^{\,2}
\end{equation}
At the corresponding power level, this equivalence becomes
\begin{equation}
M \dot{\phi} \ddot{\phi}
\quad\longleftrightarrow\quad
m\dot{x}\ddot{x}
\quad\longleftrightarrow\quad
L\dot {Q} \,\ddot {Q}.
\end{equation}
Thus, $M$ plays the role of an effective thermodynamic inertial coefficient,
analogous to the mechanical mass $m$ and the electrical inductance $L$.

\subsection{Pressure--Volume Contribution:
$\dot{U}_{V}^{\rm dyn}=-\delta p\,\dot V$}

We now consider the pressure--volume contribution to the rate of
change of the internal energy,
\begin{equation}
\dot {U}_{V}^{dyn}=-p \dot{V}.
\label{eq:pV_original}
\end{equation}
In contrast to treating the pressure as a fixed parameter, we take into
account its thermodynamic response to a small change of volume about a
mechanically stable equilibrium state. Let $V_0$ and $p_0$ denote the
equilibrium volume and pressure, respectively, and define
\begin{equation}
\delta V \equiv V - V_{0},
\qquad
\delta p \equiv p - p_{0}.
\label{eq:pV_deviations}
\end{equation}
For sufficiently small deviations from equilibrium, the pressure can be
expanded to first order in the volume displacement. For the thermodynamic
representation $U=U(S,V,N)$, and considering the volume response at fixed
entropy and particle number, one has
\begin{equation}
p = p_{0} +
\left(
\frac{\partial p}{\partial V}
\right)_{S,N;0}
\delta V + \mathcal{O}(\delta V^{2}).
\label{eq:p_expansion}
\end{equation}
Since pressure and volume changes  at inverse ratio, one can write,
\begin{equation}
\left(
\frac{\partial p}{\partial V}
\right)_{S,N;0}<0.
\label{eq:mechanical_stability}
\end{equation}
It is therefore convenient to introduce the positive volume-response
coefficient
\begin{equation}
\Gamma_{V} \equiv - \left( \frac{\partial p}{\partial V} \right)_{S,N;0} > 0.
\label{eq:GammaV}
\end{equation}
The pressure response is then
\begin{equation}
\delta p = - \Gamma_{V} \,\delta V.
\label{eq:pressure_volume_response}
\end{equation}
Eq.~(\ref{eq:pressure_volume_response}) is the thermodynamic origin
of the restoring character of the pressure--volume channel. An expansion
of the system increases its volume, $\delta V>0$, but produces a negative
pressure deviation, $\delta p<0$, whereas a compression produces the
opposite response. Thus, close to a stable equilibrium state, the
pressure response opposes the volume displacement.

The total pressure--volume power can now be separated into an equilibrium
background contribution and a contribution generated by the deviation
from equilibrium,
\begin{equation}
 p\dot{V} =  p_{0} \dot {V} + \delta p \,\dot{V}.
\label{eq:pV_split}
\end{equation}
The first term represents the work rate associated with the equilibrium
background pressure. The part relevant to the restoring dynamics is
therefore defined as
\begin{equation}
\dot {U}_{V}^{\rm dyn} \equiv -\delta p\,\dot {V}.
\label{eq:UV_dynamic}
\end{equation}
Using Eq.~(\ref{eq:pressure_volume_response}), this becomes
\begin{equation}
\dot {U}_{V}^{\rm dyn} = \Gamma_{V} \delta {V} \,\dot {V}.
\label{eq:UV_deltaV}
\end{equation}

To express this contribution in terms of the generalized transfer
coordinate $\phi$, we assume that the thermodynamic mode under
consideration can be parametrized locally by $\phi$. To leading order
near equilibrium, the volume displacement along this mode can then be
written as
\begin{equation}
\delta V=\lambda_{V} \phi,
\label{eq:deltaV_phi}
\end{equation}
where $\lambda_V$ specifies how a displacement in the generalized
coordinate changes the physical volume. Consequently,
\begin{equation}
\dot {V} = \lambda_{V} \dot {\phi}.
\label{eq:Vdot_phi}
\end{equation}
Substitution into Eq.~(\ref{eq:UV_deltaV}) gives
\begin{equation}
\dot {U}_{V}^{\rm dyn} =
\Gamma_{V} \lambda_{V}^{2}
\phi\dot\phi.
\end{equation}
Defining the effective restoring coefficient
\begin{equation}
K \equiv \Gamma_{V} \lambda_{V}^{2},
\label{eq:K_definition}
\end{equation}
we obtain
\begin{equation}
\dot {U}_{V}^{\rm dyn} = K \phi \dot{\phi}.
\label{eq:UV_Kphi}
\end{equation}

This result is important because the term $K\phi\dot {\phi}$ has not been
introduced phenomenologically. It follows from the linear thermodynamic
response of the pressure to a volume displacement about a mechanically
stable equilibrium state, together with the local parametrization of that
thermodynamic mode by the generalized coordinate $\phi$.

If $K$ can be regarded as constant within the linear-response regime,
Eq.~(\ref{eq:UV_Kphi}) can be written as
\begin{equation}
\dot {U}_{V}^{\rm dyn} = \frac{d}{dt}
\left( \frac{1}{2}K\phi^{2} \right).
\label{eq:stored_energy_rate}
\end{equation}
Hence the pressure--volume response defines an effective quadratic stored
energy,
\begin{equation}
U_{V}^{\rm dyn} = \frac{1}{2}K\phi^{2},
\label{eq:stored_energy}
\end{equation}
up to an additive constant. At the power level this may equivalently be
written as
\begin{equation}
\dot{U}_{V}^{\rm dyn}
=
\dot\phi\,(K\phi),
\end{equation}
which identifies $K \phi$ as the generalized restoring contribution conjugate to the generalized
flow $\dot{\phi}$.

The physical analogy is therefore with an energy-storage element rather
than with a dissipative process. In a mechanical oscillator, the elastic
energy stored in a spring is
\begin{equation}
U_{\rm spring} = \frac{1}{2}kx^{2},
\qquad
\dot {U}_{\rm spring} = k x \dot{x},
\end{equation}
while in an electrical circuit the energy stored in a capacitor is
\begin{equation}
U_{C}  = \frac{Q^2}{2C}, \qquad \dot {U}_{C} = \frac{Q}{C}\dot {Q}.
\end{equation}
The thermodynamic pressure--volume contribution derived above possesses
the same quadratic storage structure,
\begin{equation}
\frac{1}{2}K\phi^{2}
\quad\longleftrightarrow\quad
\frac{1}{2}kx^{2}
\quad\longleftrightarrow\quad
\frac{1}{2} C^{-1}  Q^{2}.
\end{equation}
Accordingly, the appearance of the term $K\phi$ in the reduced dynamical
description can be traced to the finite restoring response of the
thermodynamic system against volume perturbations near mechanical
equilibrium.

At the corresponding power level, this equivalence becomes
\begin{equation}
K\phi\dot{\phi}
\quad\longleftrightarrow\quad
kx\dot{x}
\quad\longleftrightarrow\quad
C^{-1} Q \dot {Q}.
\end{equation}
Thus, $K$ plays the role of an effective thermodynamic restoring
coefficient, analogous to the spring constant $k$ in the mechanical
system and the inverse capacitance $C^{-1}$ in the electrical system.
Physically, $K$ characterizes the resistance of the thermodynamic system
to a displacement of the generalized coordinate $\phi$ from its
equilibrium state and therefore determines the amount of reversible
energy stored in the corresponding pressure--volume mode.

\subsection{Particle-Transfer Contribution: $\dot{U}_{N} ^{dyn} = \mu \dot{N}$ }

We next consider irreversible particle exchange between two interacting
thermodynamic subsystems, denoted by $1$ and $2$. The total number of
particles is assumed to be conserved during the exchange process,
\begin{equation}
dN_{1} + dN_{2} = 0.
\label{eq:particle_conservation}
\end{equation}
We introduce a generalized transfer coordinate $\phi$ by defining
\begin{equation}
d \phi = dN_{2}  = -dN_{1}
\label{eq:phi_particle_transfer}
\end{equation}
Accordingly,
\begin{equation}
\dot{\phi} = \dot N_{2} = - \dot N_{1},
\label{eq:phidot_particle_transfer}
\end{equation}
so that $\dot{\phi}$ represents the net particle-transfer rate from
subsystem $1$ to subsystem $2$.

The thermodynamic driving quantity for this process is not the absolute
chemical potential of either subsystem, but their difference. We define
\begin{equation}
\Delta\mu \equiv \mu_{1}-\mu_{2}.
\label{eq:chemical_potential_difference}
\end{equation}
With the convention adopted above, $\Delta\mu>0$ drives particles from
subsystem $1$ toward subsystem $2$, so that $\dot{\phi}>0$.
The power associated with the relaxation of the chemical-potential
difference can therefore be written as
\begin{equation}
\dot {U}_{N}^{\rm dyn} = \Delta\mu\,\dot{\phi}.
\label{eq:chemical_transfer_power}
\end{equation}
This expression already has the characteristic thermodynamic
force--flux structure: $\Delta\mu$ is the generalized chemical driving
force, while $\dot{\phi}$ is the corresponding particle-transfer flux.

The constitutive relation between these two quantities follows from
Onsager's linear irreversible thermodynamics
\cite{Onsager_1931_I,Onsager_1931_II,deGroot_1984,Prigogine_1967}.
For two subsystems sufficiently close to equilibrium and maintained at
the same temperature $T$, the entropy production associated with particle
exchange may be written as
\begin{equation}
\dot {S}_{i}^{(N)} = \frac{\Delta\mu}{T}\,\dot{\phi}.
\label{eq:entropy_production_particle}
\end{equation}
The corresponding thermodynamic force and flux are therefore identified as
\begin{equation}
X_{N} = \frac{\Delta\mu}{T}, \qquad J_{N} = \dot{\phi}.
\label{eq:force_flux_particle}
\end{equation}
Near equilibrium,
\begin{equation}
\Delta\mu \rightarrow 0,
\qquad
\dot{\phi} \rightarrow 0,
\end{equation}
and Onsager's linear-response relation gives
\begin{equation}
J_{N} = L_{N} X_{N},
\label{eq:onsager_particle}
\end{equation}
where $L_N>0$ is the particle-transfer Onsager coefficient. Substituting
Eq.~(\ref{eq:force_flux_particle}) into
Eq.~(\ref{eq:onsager_particle}) yields
\begin{equation}
\dot{\phi} = L_{N} \frac{\Delta\mu}{T}.
\label{eq:phidot_onsager}
\end{equation}
This is the precise point at which the Onsager assumption enters the
derivation. The relation
\begin{equation}
\Delta\mu \propto \dot{\phi}
\end{equation}
is therefore not introduced by analogy with mechanical damping or
electrical resistance; it follows from the linear force--flux relation
valid sufficiently close to thermodynamic equilibrium.

Solving Eq.~(\ref{eq:phidot_onsager}) for the chemical-potential
difference gives
\begin{equation}
\Delta\mu = \frac{T}{L_{N}}\dot{\phi}.
\label{eq:dmu_phidot}
\end{equation}
We now define the effective thermodynamic resistance
\begin{equation}
R \equiv \frac{T}{L_{N} },
\label{eq:R_thermodynamic}
\end{equation}
so that
\begin{equation}
\Delta\mu = R \dot{\phi}.
\label{eq:chemical_resistance_relation}
\end{equation}
The coefficient $R$ therefore measures the opposition of the
particle-transfer channel to a finite particle flux: for a given
$\dot{\phi}$, a larger $R$ requires a larger chemical-potential
difference to sustain the transfer.

Substitution into Eq.~(\ref{eq:chemical_transfer_power}) gives
\begin{equation}
\dot {U}_{N}^{\rm dyn} = R \dot{\phi}^{\,2}.
\label{eq:R_power_final}
\end{equation}
Since $L_N>0$ and $T>0$, one has $R>0$, and consequently
\begin{equation}
R \dot{\phi}^{2} \geq 0.
\label{eq:positive_dissipation}
\end{equation}
The thermodynamic meaning of this positivity follows directly from the
entropy production associated with particle transfer. Using
Eq.~(\ref{eq:entropy_production_particle}), one obtains
\begin{equation}
T\dot{S}_{i}^{(N)}
=
\Delta\mu\,\dot{\phi}
=
R\dot{\phi}^{\,2}
\geq 0,
\label{eq:entropy_production_R}
\end{equation}
which is consistent with the second law of thermodynamics.

Thus, $R\dot{\phi}^{\,2}$ represents the irreversible dissipation rate
associated with relaxation of the chemical-potential difference.

In terms of power, the same expression can be factorized as
\begin{equation}
\dot {U}_{N}^{\rm dyn} = \dot{\phi} \left( R \dot{\phi} \right).
\end{equation}
Hence the generalized dissipative contribution conjugate to the
generalized flow $\dot{\phi}$ is
$R \dot{\phi}$. The physical interpretation becomes transparent when this result is
compared with mechanical and electrical dissipation. For a viscously
damped mechanical system,
\begin{equation}
P_{\rm diss}^{(m)} = b \dot{x}^{\,2},
\end{equation}
whereas Joule dissipation in an electrical resistor is
\begin{equation}
P_{\rm diss}^{(e)} = R_{\rm el} \dot{Q}^2.
\end{equation}
The particle-transfer channel has precisely the same quadratic
force--flux structure,
\begin{equation}
P_{\rm diss}^{(N)} = R\dot{\phi}^{\,2}.
\end{equation}
The correspondence at the power level is therefore
\begin{equation}
R\dot{\phi}^{\,2}
\quad\longleftrightarrow\quad
b\dot{x}^{\,2}
\quad\longleftrightarrow\quad
R_{\rm el} \dot{Q}^{2}.
\label{eq:dissipative_power_correspondence}
\end{equation}
Equivalently, at the level of the generalized dissipative force,
\begin{equation}
R \dot{\phi}
\quad\longleftrightarrow\quad
b\dot{x}
\quad\longleftrightarrow\quad
R_{\rm el} \dot{Q}.
\label{eq:dissipative_force_correspondence}
\end{equation}
Thus, the appearance of the term $R\dot{\phi}$ in the reduced
thermodynamic dynamics is not postulated from the mechanical or
electrical analogy. It follows from Onsager's linear force--flux relation
for irreversible particle transfer. The analogy emerges only after this
thermodynamic result has been established: $R$ acts as an effective
resistance to particle transfer, in the same structural sense that
viscous friction opposes mechanical motion and electrical resistance
opposes electric current.

\subsection{Combined Dynamical Energy Balance}

The three thermodynamic contributions derived above are formulated within
the same near-equilibrium linear-response framework. The apparently
different roles of temperature, pressure, and chemical potential do not
represent an inconsistency, but reflect the distinction between
reference-state quantities and dynamical deviations from equilibrium.
The temperature is evaluated at its reference value,
$T\simeq T_{0}$, while the pressure deviation
$\delta p=p-p_{0}$ and the chemical-potential difference
$\Delta\mu=\mu_{1} - \mu_{2}$ are retained as dynamical quantities because
they provide the leading restoring and dissipative responses,
respectively. The corresponding response coefficients are likewise
evaluated at the reference state and treated as constant to leading
order within the linear-response regime.

The three dynamical energy contributions can therefore be summarized as
\begin{equation}
\dot{U}_{S}^{dyn} =  T\frac{d}{dt}(S_{\rm eq}-S) = M\dot{\phi} \,\ddot{\phi},
\label{eq:three_mapping_M}
\end{equation}
\begin{equation}
\dot{U}_{N}^{dyn} =
\Delta\mu\,\dot{\phi} = R \dot{\phi}^{\,2},
\label{eq:three_mapping_R}
\end{equation}
\begin{equation}
\dot{U}_{V}^{dyn} = - (p - p_{0} ) \dot{V} = K \phi  \dot{\phi}.
\label{eq:three_mapping_K}
\end{equation}
Returning now to the dynamical energy balance introduced in
Eq.~(\ref{eq:fundamental_power_balance}), the total dynamical energy-transfer rate is
therefore
\begin{equation}
\dot {U}_{\rm dyn} = M \dot{\phi} \,\ddot{\phi} + R \dot{\phi}^{\,2} + K \phi \dot{\phi}.
\label{eq:combined_dynamic_power}
\end{equation}
Every term contains the common generalized flux $\dot{\phi}$. Therefore, we can write
\begin{equation}
\dot {U}_{\rm dyn} =
\dot{\phi} \left( M \ddot{\phi} + R \dot{\phi} + K \phi \right).
\end{equation}
For $\dot{\phi} \neq 0$, division by the generalized flux gives
\begin{equation}
M \ddot{\phi} + R \dot{\phi} + K \phi = \frac{\dot U_{\rm dyn}}{\dot{\phi} }.
\label{eq:generalized_thermodynamic_equation}
\end{equation}
Defining the generalized thermodynamic driving potential as
\begin{equation}
\mathcal{A}(t) \equiv
\frac{\dot {U}_{\rm dyn}}{\dot{\phi} },
\label{eq:generalized_affinity}
\end{equation}
the final equation takes the compact form
\begin{equation}
M \ddot{\phi} + R \dot{\phi}  + K \phi = \mathcal{A}(t).
\label{eq:final_MRK_equation}
\end{equation}
The resulting second-order equation therefore combines three physically
distinct thermodynamic mechanisms within a single dynamical structure:
the nonequilibrium entropy deficit provides the inertial-like term
$M\ddot{\phi}$, irreversible particle transfer provides the dissipative
term $R\dot{\phi}$, and the pressure response around mechanical
equilibrium provides the restoring term $K\phi$. Thus, the
inertial--dissipative--restoring structure is not imposed by analogy
with a mechanical oscillator or an electrical circuit, but emerges from
the corresponding near-equilibrium thermodynamic responses.

To test the consistency of the solutions obtained in Eqs.~(\ref{eq:combined_dynamic_power}) and (\ref{eq:final_MRK_equation}), we performed a detailed dimensional analysis of all terms entering the corresponding dynamical formulations. The analysis shows that the dimensions of the derived quantities and effective coefficients are mutually consistent and that each term in the resulting equations has the required physical dimensions. The details of the dimensional analysis for $U$ and $\dot{U}$ are presented in Appendices~\ref{dim-an-U} and \ref{dim-an-dotU}, respectively. These results provide an independent consistency check supporting the dimensional validity of the derived dynamical equations.

The result given in Eq.~(\ref{eq:final_MRK_equation}) represents the equation of motion of an open thermodynamic system capable of exchanging heat and particles with its surroundings. This equation demonstrates that the thermodynamic system is dynamically equivalent to a mechanical system described by a forced damped harmonic oscillator and to an electrical system represented by a driven RLC circuit supplied by an external voltage source. As will be seen in the Table~\ref{tab:dynamical_correspondence}, these three physically distinct systems share the same universal dynamical structure.

The dimensionless form of Eqs.~(\ref{eq:final_MRK_equation})  is given by Eq.~(\ref{undim_ther}). This dimensionless equation is equivalent to Eq.~(\ref{undim_mec}) for the mechanical system and Eq.~(\ref{undim_elec}) for the electrical system. Consequently, the dimensionless dynamics of all three systems can be expressed in the common form given by Eq.~(\ref{undim_com}).

\begin{table*}[t]
\centering
\caption{Dynamical correspondence among mechanical, electrical, and
thermodynamic systems.}
\label{tab:dynamical_correspondence}
\begin{tabular}{lccc}
\hline\hline
Dynamical quantity
& Mechanical system & Electrical system & Thermodynamic system
\\
\hline

Generalized coordinate & $x$  & $q$    &  $\phi$
\\[1.5mm]

Generalized rate  &  $\dot{x}=v$  &  $\dot{q}=I$
&
$\dot{\phi}$
\\[1.5mm]

Generalized acceleration
&
$\ddot{x}=\dot{v}$
&
$\ddot{q}=\dot{I}$
&
$\ddot{\phi}$
\\[1.5mm]

Inertial coefficient
&
$m$
&
$L$
&
$M$
\\[1.5mm]

Dissipative coefficient
&
$b$
&
$R _{\rm el}$
&
$R$
\\[1.5mm]

Restoring coefficient
&
$k$
&
$\dfrac{1}{C}$
&
$K$
\\[2mm]

Driving/source term
&
$F(t)$
&
$V(t)$
&
$\dot{U}$
\\[2mm]

Equation of motion
&
$\displaystyle
m\ddot{x}+b\dot{x}+kx=F(t)$
&
$\displaystyle
L\ddot{q}+R_{\rm el} \dot{q}+\frac{1}{C}q=V(t)$
&
$\displaystyle
M\ddot{\phi}+R \dot{\phi}+K\phi=\dot{U}$
\\[4mm]

Flux form
&
$\displaystyle
m\dot{v}+bv+kx=F(t)$
&
$\displaystyle
L\dot{I}+R_{\rm el} I+\frac{1}{C}q=V(t)$
&
$\displaystyle
M\dot{\psi}+R \psi+K\phi=\dot{U}$
\\[3mm]

Kinematic relation
&
$\dot{x}=v$
&
$\dot{q}=I$
&
$\dot{\phi}=\psi$
\\

Power / energy-transfer rate
&
$\displaystyle
P_{\rm mec}=F(t)\dot{x}$
&
$\displaystyle
P_{\rm elec}=V(t)I=V(t)\dot{q}$
&
$\displaystyle
P_{\rm th}=\dot{U}$
\\[3mm]

Source-term dimension
&
$\displaystyle [F]=\mathrm{N}$
&
$\displaystyle [V]=\mathrm{V}$
&
$\displaystyle [\dot{U}]=\mathrm{W}$
\\[3mm]

\hline\hline
\end{tabular}
\end{table*}


\section{Lagrange Formulation of the Thermodynamic System} \label{Lagrange}

We first consider the nondissipative limit,
\begin{equation}
R = 0.
\end{equation}

From Eq.(\ref{eq:final_MRK_equation}),  the homogeneous thermodynamic equation of motion is then written as
\begin{equation}
M \ddot{\phi} + K \phi = 0.
\label{eq:ther_motion_homogeneous}
\end{equation}
Multiplying Eq.~(\ref{eq:ther_motion_homogeneous}) by $\dot{\phi}$ gives
\begin{equation}
M \ddot{\phi} \dot{\phi} + K \phi\dot{\phi} = 0.
\end{equation}
Using
\begin{equation}
M \ddot{\phi} \dot{\phi} = \frac{d}{dt} \left( \frac{1}{2} M \dot{\phi}^{\,2} \right),
\end{equation}
and
\begin{equation}
K \phi\dot{\phi} = \frac{d}{dt} \left( \frac{1}{2}K\phi^{2} \right),
\end{equation}
we obtain
\begin{equation}
\frac{d}{dt} \left[ \frac{1}{2}M\dot{\phi}^{\,2} + \frac{1}{2}K\phi^{2} \right] = 0.
\label{eq:ther_energy_conservation}
\end{equation}
Thus, the conserved effective energy can be written as
\begin{equation}
E_{\rm th} = T_{\rm th} + V_{\rm th},
\end{equation}
where
\begin{equation}
T_{\rm th} = \frac{1}{2}M\dot{\phi}^{\,2}
\label{eq:ther_kinetic_energy}
\end{equation}
is the kinetic-like contribution and
\begin{equation}
V_{\rm th} = \frac{1}{2}K\phi^{2}
\label{eq:ther_potential_energy}
\end{equation}
is the potential-like contribution. Therefore,
\begin{equation}
E_{\rm th} = \frac{1}{2}M\dot{\phi}^{\,2} + \frac{1}{2}K\phi^{2}.
\label{eq:ther_total_energy}
\end{equation}

The corresponding Lagrangian is defined in the standard form
\begin{equation}
\mathcal{L}_{\rm th} = T_{\rm th} - V_{\rm th},
\end{equation}
which gives
\begin{equation}
\mathcal{L}_{\rm th} = \frac{1}{2}M\dot{\phi}^{\,2} - \frac{1}{2}K\phi^{2}.
\label{eq:ther_lagrangian}
\end{equation}

The Euler--Lagrange equation is
\begin{equation}
\frac{d}{dt} \left( \frac{\partial \mathcal{L}_{\rm th}} {\partial \dot{\phi}} \right) - \frac{\partial \mathcal{L}_{\rm th}}
{\partial \phi} = 0.
\label{eq:ther_euler_lagrange}
\end{equation}
From Eq.~(\ref{eq:ther_lagrangian}),
\begin{equation}
\frac{\partial \mathcal{L}_{\rm th}}
{\partial \dot{\phi}} = M \dot{\phi},
\end{equation}
and hence
\begin{equation}
\frac{d}{dt} \left( \frac{\partial \mathcal{L}_{\rm th}} {\partial \dot{\phi}} \right) = M \ddot{\phi}.
\end{equation}
Furthermore,
\begin{equation}
\frac{\partial \mathcal{L}_{\rm th}} {\partial \phi} = - K\phi.
\end{equation}
Substitution into Eq.~(\ref{eq:ther_euler_lagrange}) yields
\begin{equation}
M \ddot{\phi} + K \phi = 0,
\end{equation}
which reproduces the homogeneous thermodynamic equation of motion.

The canonical momentum conjugate to $\phi$ is defined by
\begin{equation}
p_{\phi} = \frac{\partial \mathcal{L}_{\rm th}} {\partial \dot{\phi}},
\end{equation}
and therefore
\begin{equation}
p_{\phi} = M \dot{\phi}.
\label{eq:ther_canonical_momentum}
\end{equation}
Thus, we obtain the corresponding canonical pair as
\begin{equation}
(\phi,p_{\phi}).
\end{equation}

On the other hand, the Hamiltonian is obtained through the Legendre transformation
\begin{equation}
\mathcal{H}_{\rm th} = p_{\phi}\dot{\phi} - \mathcal{L}_{\rm th}.
\end{equation}
Using
\begin{equation}
\dot{\phi} = \frac{p_{\phi}}{M},
\end{equation}
we obtain
\begin{equation}
\mathcal{H}_{\rm th} = \frac{p_{\phi}^{2}}{2M} + \frac{1}{2}K\phi^{2}.
\label{eq:ther_hamiltonian}
\end{equation}

The dynamical correspondence among the three systems can also be examined
at the level of their energy-related quantities. As summarized in
Table~\ref{tab:energy_correspondence}, the inertial terms give rise to
kinetic-like contributions proportional to the square of the corresponding
generalized velocities, whereas the restoring terms generate potential-like
contributions quadratic in the generalized coordinates. In the mechanical
system, these contributions correspond to the kinetic energy of the mass
and the elastic potential energy stored in the spring. In the electrical
system, their counterparts are the magnetic energy stored in the inductor
and the electric energy stored in the capacitor. The thermodynamic system
possesses formally analogous terms, represented by
$\frac{1}{2}M\dot{\phi}^{\,2}$ and $\frac{1}{2}K\phi^{2}$.

The correspondence extends naturally to the dissipative and driving
contributions. The terms $b\dot{x}^{\,2}$, $R\dot{q}^{\,2}$, and
$R\dot{\phi}^{\,2}$ represent the corresponding dissipative contributions,
while multiplication of each equation of motion by its generalized
velocity produces the associated generalized power-input term. Thus, the
energy-level representation preserves the same mathematical correspondence
already identified at the level of the equations of motion. This provides
a further indication that the mechanical, electrical, and thermodynamic
systems share a common underlying dynamical structure.

\begin{table*}[t]
\centering
\caption{Energy-related correspondence among mechanical, electrical,
and thermodynamic dynamical systems.}
\label{tab:energy_correspondence}
\begin{tabular}{lccc}
\hline\hline
Energy-related quantity  & Mechanical system &  Electrical system & Thermodynamic system
\\
\hline

Kinetic-like term
&
$\displaystyle \frac{1}{2}m\dot{x}^{\,2}$
&
$\displaystyle \frac{1}{2}L\dot{q}^{\,2}
=
\frac{1}{2}LI^2$
&
$\displaystyle \frac{1}{2}M\dot{\phi}^{\,2}$
\\[3mm]

Potential-like term
&
$\displaystyle \frac{1}{2}kx^2$
&
$\displaystyle \frac{q^2}{2C}$
&
$\displaystyle \frac{1}{2}K\phi^2$
\\[3mm]

Dissipative power
&
$\displaystyle b\dot{x}^{\,2}$
&
$\displaystyle R_{\rm el} I^2
=
R_{\rm el} \dot{q}^{\,2}$
&
$\displaystyle R\dot{\phi}^{\,2}$
\\[3mm]

External power input
&
$\displaystyle F(t)\dot{x}$
&
$\displaystyle V(t) \dot{q}$
&
$\displaystyle \dot{U}\,\dot{\phi}$
\\

\hline\hline
\end{tabular}
\end{table*}

Having established the Lagrangian formulation of the thermodynamic
system, we can now compare its canonical structure with those of the
corresponding mechanical and electrical systems. For this purpose, we
consider the nondissipative limit, $R=0$, so that all three systems can
be described within the standard conservative Lagrangian framework,
$\mathcal{L}=T-V$. Table~\ref{tab:canonical-correspondence} summarizes
the resulting correspondence. The generalized coordinates
$x$, $q$, and $\phi$, together with their corresponding generalized
velocities and canonical momenta, play analogous roles in the three
systems. In particular, the canonical pairs $(x,p_x)$, $(q,p_q)$, and
$(\phi,p_\phi)$ exhibit the same mathematical structure. This
correspondence shows that, in the nondissipative limit, the effective
thermodynamic dynamics admits the same canonical formulation as the
mechanical oscillator and the electrical LC circuit.

\begin{table*}[t]
\centering
\caption{Canonical correspondence among mechanical, electrical,
and thermodynamic dynamical systems.}
\label{tab:canonical-correspondence}
\begin{tabular}{lccc}
\hline\hline
Canonical quantity
&
Mechanical system
&
Electrical system
&
Thermodynamic system
\\
\hline

Generalized coordinate
&
$x$
&
$q$
&
$\phi$
\\[2mm]

Generalized velocity
&
$\dot{x}$
&
$\dot{q}=I$
&
$\dot{\phi}$
\\[2mm]

Lagrangian
&
$\displaystyle
\mathcal{L} = \frac{1}{2}m\dot{x}^{2} - \frac{1}{2}kx^{2}$
&
$\displaystyle
\mathcal{L} = \frac{1}{2}L\dot{q}^{\,2} - \frac{q^{2}}{2C}$
&
$\displaystyle
\mathcal{L}
=
\frac{1}{2}M\dot{\phi}^{\,2}
-
\frac{1}{2}K\phi^{2}$
\\[4mm]

Canonical momentum
&
$\displaystyle
p_{x} = \frac{\partial\mathcal{L}}{\partial\dot{x}} = m\dot{x}$
&
$\displaystyle
p_{q} = \frac{\partial\mathcal{L}}{\partial\dot{q}} = L\dot{q} = LI$
&
$\displaystyle
p_{\phi} = \frac{\partial\mathcal{L}}{\partial\dot{\phi}}
=
M\dot{\phi}$
\\[4mm]

Canonical pair &  $(x,p_{x})$   &    $(q,p_{q})$    &    $(\phi,p_{\phi})$
\\

\hline\hline
\end{tabular}
\end{table*}

\section{Discussion and Conclusion} \label{Con}

The central result of the present work is that a mechanical-type equation of motion can be obtained directly from the first law of thermodynamics for an open system through dimensional reduction. This point is essential for interpreting the present formulation. The inertial--dissipative--restoring structure of the resulting equation is not introduced phenomenologically, nor is it postulated by analogy with a mechanical oscillator or an electrical circuit. Instead, its three dynamical contributions originate from distinct thermodynamic mechanisms: the nonequilibrium entropy deficit generates the inertial-like term $M\ddot{\phi}$, irreversible particle transfer produces the dissipative term $R\dot{\phi}$, and the pressure response about mechanical equilibrium gives rise to the restoring term $K\phi$. The oscillator-like structure therefore emerges from the thermodynamic description itself.

This result changes the role of the mechanical and electrical analogies in the present problem. Conventionally, an analogy begins by identifying variables or equations in two different physical systems that exhibit similar mathematical behavior. Here, however, no oscillator equation is assumed at the outset. The starting point is the first law of thermodynamics, and only after its reduction to a second-order dynamical equation does the correspondence with the forced damped harmonic oscillator and the driven RLC circuit become apparent. The mechanical--electrical--thermodynamic correspondence is therefore a consequence of the derivation rather than an assumption used to construct it.

The dimensional structure of the formulation provides an important consistency test of this procedure. Detailed dimensional analyses of Eqs.~(\ref{eq:combined_dynamic_power}) and (\ref{eq:final_MRK_equation}) show that the effective coefficients and dynamical variables combine consistently and that all terms entering each equation possess the required physical dimensions. The corresponding analyses for $U$ and $\dot{U}$ are given in Appendices~\ref{dim-an-U} and \ref{dim-an-dotU}. Although dimensional consistency alone does not establish the physical validity of a dynamical theory, it provides a nontrivial independent check that the dimensional reduction leading from the thermodynamic relation to the resulting equation of motion is internally consistent.

The equivalence of the three systems is demonstrated in Appendix~\ref{dim-all-sys}.
The thermodynamic equation, Eq.~(\ref{eq:final_MRK_equation}), reduces to the dimensionless form given in Eq.~(\ref{undim_ther}). The mechanical oscillator and the RLC circuit similarly reduce to Eqs.~(\ref{undim_mec}) and (\ref{undim_elec}), respectively. All three can then be represented by the common dimensionless equation given in Eq.~(\ref{undim_com}). Thus, despite the different physical meanings and dimensions of their original variables and parameters, the three systems share the same reduced dynamical structure. The correspondence summarized in Table~\ref{tab:dynamical_correspondence} should therefore be understood at the level of dynamical structure rather than as a literal identification of thermodynamic, mechanical, and electrical quantities.

This distinction is physically important. Mass, resistance, inductance, entropy, particle number, and volume are clearly not interchangeable physical quantities. What is common among the three systems is instead the organization of their dynamics into inertial, dissipative, restoring, and driving contributions. The coefficients $M$, $R$, and $K$ introduced by the thermodynamic reduction should therefore be interpreted as effective dynamical coefficients characterizing these roles within thermodynamic evolution. Their significance lies not in reproducing mechanical parameters microscopically, but in revealing that thermodynamic evolution can possess the same mathematical organization as canonical second-order dynamical systems.

The present formulation consequently suggests a broader interpretation of the first law. In its conventional differential form, the first law specifies the relation among changes in internal energy, entropy, volume, and particle number. The dimensional reduction developed here shows that, under the assumptions employed in the present treatment, these thermodynamic contributions can also be reorganized into a dynamical equation governing the evolution of a generalized thermodynamic coordinate. In this sense, the first law contains sufficient structure to generate a mechanical-type dynamical representation once the relevant near-equilibrium responses are identified.

The main conceptual advance is therefore not the observation that oscillator equations occur in different areas of physics; such mathematical analogies are well known. Rather, it is that the oscillator-type equation is obtained here from the thermodynamic law itself. The subsequent equivalence with the mechanical oscillator and the RLC circuit emerges only after this thermodynamic derivation has been completed. This provides a direct bridge between thermodynamic evolution and classical dynamical-systems theory and places the three physically distinct systems within a common dynamical class.

The present derivation relies on the near-equilibrium response relations and on the dimensional reduction introduced above. Accordingly, the resulting equation should not be interpreted as a universal microscopic equation of motion for arbitrary thermodynamic systems far from equilibrium. Rather, it establishes that, within the regime considered here, the first law admits a systematic reduction to a second-order dynamical form. Determining how far this structure survives beyond the present approximations, how the effective coefficients depend on microscopic dynamics, and whether nonlinear or strongly nonequilibrium extensions generate broader classes of dynamical equations are natural questions for future investigation.

\section*{ACKNOWLEDGMENTS} 

The author would like to thank Güngör Gündüz for identifying several typographical errors and for his valuable comments and suggestions.

\appendix  \label{appendix}


\section{Dimensional Analysis of the First Law}  \label{dim-an-U}

We first consider the dimensions of the internal energy. Since
\begin{equation}
[U]=\mathrm{J},
\end{equation}
the time rate of change of the internal energy has the dimension
\begin{equation}
[\dot{U}] = \frac{[dU]}{[dt]} = \mathrm{J\,s^{-1}} = \mathrm{W}.
\end{equation}
For the entropy contribution, we have
\begin{equation}
[T]=\mathrm{K}, \qquad [S]=\mathrm{J\,K^{-1}},
\end{equation}
and therefore
\begin{equation}
[T\dot{S}] = [T] \frac{[dS]}{[dt]}
= \mathrm{K} \left( \mathrm{J\,K^{-1}\,s^{-1}} \right) = \mathrm{J\,s^{-1}} = \mathrm{W}.
\end{equation}
Similarly, for the pressure--volume contribution,
\begin{equation}
[p] = \mathrm{Pa} = \mathrm{N\,m^{-2}} = \mathrm{J\,m^{-3}},
\end{equation}
while
\begin{equation}
[\dot{V}] = \mathrm{m^{3}\,s^{-1}}.
\end{equation}
Hence,
\begin{equation}
[p\dot{V}] = [p][\dot{V}] = \left( \mathrm{J\,m^{-3}} \right) \left( \mathrm{m^{3}\,s^{-1}} \right) = \mathrm{J\,s^{-1}} =
\mathrm{W}.
\end{equation}
Finally, if $N$ denotes the amount of substance measured in moles, the
chemical potential has the unit
\begin{equation}
[\mu] = \mathrm{J\,mol^{-1}},
\end{equation}
whereas the particle-transfer rate has the unit
\begin{equation}
[\dot{N}] = \mathrm{mol\,s^{-1}}.
\end{equation}
It therefore follows that
\begin{equation}
[\mu\dot{N}] = [\mu][\dot{N}] = \left( \mathrm{J\,mol^{-1}} \right)
\left( \mathrm{mol\,s^{-1}} \right) = \mathrm{J\,s^{-1}} = \mathrm{W}.
\end{equation}
Consequently, all terms appearing in the time-dependent form of the
first law have the same physical dimension,
\begin{equation}
[\dot{U}] = [T\dot{S}] = [p\dot{V}] = [\mu\dot{N}] = \mathrm{W}.
\end{equation}
In terms of the fundamental SI units, the watt is given by
\begin{equation}
\mathrm{W} = \mathrm{J\,s^{-1}} = \mathrm{kg\,m^{2}\,s^{-3}}.
\end{equation}
Thus, each term in the time-dependent thermodynamic energy balance
represents a rate of energy transfer, or equivalently, a power.

\section{Dimensional Consistency of the Generalized Thermodynamic Equation}  \label{dim-an-dotU}

Before considering the dynamical form of the first law, we first
establish the dimensions of the internal energy $U$ and its time
derivative $\dot{U}$.
The internal energy has the dimension of energy,
\begin{equation}
[U] = M L^{2} T^{-2} = \mathrm{J}.
\end{equation}
Taking the time derivative gives
\begin{equation}
[\dot{U}] = \frac{\left[ dU\right] }{\left[dt \right]} = M L^{2} T^{-3} = \mathrm{J\,s^{-1}} = \mathrm{W}.
\end{equation}
Thus, while $U$ represents energy, $\dot{U}$ represents power.
Consequently, every term contributing to $\dot{U}$ must have the
dimension of power,
\begin{equation}
[\dot{U}] = [T\dot{S}] = [p\dot{V}] = [\mu\dot{N}] = \mathrm{W}.
\end{equation}

The dimensional ratio between the internal-energy rate $\dot{U}$ and
the generalized transfer rate $\dot{\phi}$ is
\begin{equation}
\frac{ \left[ \dot{U}\right]}{ \left[\dot{\phi}\right]} = \frac{ \mathrm{J\,s^{-1}} }{ \mathrm{mol\,s^{-1}} } = \mathrm{J\,mol^{-1}}.
\end{equation}
Since
\begin{equation}
[\phi]=\mathrm{mol}, \qquad [U]=\mathrm{J},
\end{equation}
we equivalently obtain
\begin{equation}
\frac{ \left[ \dot{U}\right]}{ \left[\dot{\phi}\right]} = \frac{[U]}{[\phi]} = \mathrm{J\,mol^{-1}}.
\end{equation}
Thus, $\dot{U}/\dot{\phi}$ has the dimension of energy per mole,
which is precisely the dimension of a chemical potential or,
more generally, a thermodynamic driving potential conjugate to the
transferred amount of matter.

\subsection{Entropy Contribution and the Coefficient $M$}

The first contribution is written in terms of the entropy deficit
relative to equilibrium,
\begin{equation}
\Delta S \equiv S_{\rm eq} - S.
\label{eq:entropy_deficit_dimension}
\end{equation}
The corresponding relation derived above is
\begin{equation}
T\frac{d}{dt}(S_{\rm eq} - S) = M \dot{\phi} \,\ddot{\phi}.
\label{eq:entropy_inertial_relation}
\end{equation}

The thermodynamic temperature has the dimension
$\left[ T  \right]=\mathrm{K}$
whereas entropy has the dimension
$\left[ S \right ] = [\Delta S]= \mathrm{J\,K^{-1}}$. 
Consequently,
\begin{equation}
\left[ T\right]   \frac{ \left[  d\Delta S \right] }{ \left[ dt \right]}  = \mathrm{K}
\frac{\mathrm{J\,K^{-1}}}{\mathrm{s}} = \mathrm{J\,s^{-1}}
\label{eq:entropy_power_dimension}
\end{equation}
Thus, the entropy contribution indeed has the dimension of power.
On the right-hand side of
Eq.~(\ref{eq:entropy_inertial_relation}),
\begin{equation}
\left[ \dot{\phi} \,\ddot {\phi} \right] =
\left[ \dot{\phi} ][\ddot{\phi} \right] =
\left( \mathrm{mol\,s^{-1}} \right)
\left( \mathrm{mol\,s^{-2}} \right) = \mathrm{mol^2\,s^{-3}}.
\end{equation}
Therefore, dimensional consistency requires
\begin{equation}
[M]\, \mathrm{mol^2\,s^{-3}} = \mathrm{J\,s^{-1}}.
\end{equation}
Solving for the dimension of $M$ gives
\begin{equation}
[ M ] = \frac{\mathrm{J\,s^{-1}}} {\mathrm{mol^2\,s^{-3}}} = \mathrm{J\,s^2\,mol^{-2}}
\label{eq:M_dimension}
\end{equation}
This result can be checked independently from the extended nonequilibrium entropy relation
\begin{equation}
S_{\rm eq}-S = \frac{\alpha}{2}\dot{\phi}^{\,2}.
\label{eq:entropy_flux_relation_dimension}
\end{equation}
Since
\begin{equation}
[S_{\rm eq}-S] = \mathrm{J\,K^{-1}},
\end{equation}
and
\begin{equation}
[\dot{\phi}^2] = \mathrm{mol^2\,s^{-2}},
\end{equation}
the coefficient $\alpha$ must have the dimension
\begin{equation}
[\alpha] = \frac{[S_{\rm eq}-S]}{[\dot{\phi} ]^2} = \frac{\mathrm{J\,K^{-1}}} {\mathrm{mol^2\,s^{-2}}} = \mathrm{J\,s^2\,K^{-1}\,mol^{-2}}.
\label{eq:alpha_dimension}
\end{equation}
Using the identification
\begin{equation}
M=\alpha T,
\label{eq:M_alphaT_dimension}
\end{equation}
we obtain
\begin{equation}
[M] = [\alpha][T] = \left( \mathrm{J\,s^2\,K^{-1}\,mol^{-2}} \right)
\left( \mathrm{K} \right) = \mathrm{J\,s^2\,mol^{-2}}
\end{equation}
which is exactly the same result as
Eq.~(\ref{eq:M_dimension}).

Finally, we obtain
\begin{equation}
[M\dot{\phi}\,\ddot{\phi}  ] = \left( \mathrm{J\,s^2\,mol^{-2}} \right)
\left( \mathrm{mol\,s^{-1}} \right)
\left( \mathrm{mol\,s^{-2}} \right) = \mathrm{J\,s^{-1}}
\end{equation}
Hence, the inertial-like entropy contribution is dimensionally
consistent with an energy-transfer rate.

\subsection{Pressure--Volume Contribution and the Coefficient $K$}

The restoring contribution obtained from the pressure--volume work is
\begin{equation}
- ( p - p_{0} ) \dot {V} = K \phi \dot{\phi}.
\label{eq:pressure_restoring_dimension}
\end{equation}
Pressure has the SI dimension
\begin{equation}
 \left[ p  \right] = \mathrm{Pa} = \mathrm{N\,m^{-2}} = \mathrm{J\,m^{-3}}.
\label{eq:pressure_dimension}
\end{equation}
The volume has the dimension
\begin{equation}
[V]=\mathrm{m^3},
\end{equation}
and therefore
\begin{equation}
[\dot V] = \mathrm{m^3\,s^{-1}}.
\end{equation}
Consequently,
\begin{equation}
[(p - p_{0} )\dot {V} ] = [p][ \dot{V} ] = \left( \mathrm{J\,m^{-3}} \right)
\left( \mathrm{m^3\,s^{-1}} \right) = \mathrm{J\,s^{-1}}.
\label{eq:pressure_volume_power_dimension}
\end{equation}
Thus, the pressure--volume contribution has the dimension of power.

On the right-hand side of
Eq.~(\ref{eq:pressure_restoring_dimension}),
\begin{equation}
[  \phi  \dot{\phi} ] = [ \phi ][\dot {\phi} ] = \left( \mathrm{mol} \right)
\left( \mathrm{mol\,s^{-1}} \right) = \mathrm{mol^2\,s^{-1}}.
\end{equation}
Dimensional consistency therefore requires
\begin{equation}
[K]\, \mathrm{mol^2\,s^{-1}} = \mathrm{J\,s^{-1}},
\end{equation}
which gives
\begin{equation}
[K] = \frac{\mathrm{J\,s^{-1}}} {\mathrm{mol^2\,s^{-1}}} = \mathrm{J\,mol^{-2}}.
\label{eq:K_dimension}
\end{equation}
This result can also be checked independently from the relation
\begin{equation}
V - V_{0}=\lambda \phi .
\label{eq:volume_q_dimension}
\end{equation}
Since
\begin{equation}
[V - V_{0}] = \mathrm{m^3}
\end{equation}
and
\begin{equation}
[ \phi ]=\mathrm{mol},
\end{equation}
the conversion coefficient $\lambda$ has the dimension
\begin{equation}
[\lambda] = \mathrm{m^3\,mol^{-1}}.
\label{eq:lambda_dimension}
\end{equation}
The pressure response coefficient was defined as
\begin{equation}
\kappa_{V} = - \left( \frac{\partial p}{\partial V} \right)_{0}.
\end{equation}
Its dimension is therefore
\begin{equation}
[\kappa_{V}] = \frac{[p]}{[V]} = \frac{\mathrm{J\,m^{-3}}} {\mathrm{m^3}} = \mathrm{J\,m^{-6}}.
\label{eq:kappa_dimension}
\end{equation}
Since the effective restoring coefficient is
\begin{equation}
K=\kappa_V\lambda^2,
\end{equation}
we obtain
\begin{equation}
[K] = \mathrm{J\,mol^{-2}}. 
\end{equation}
in exact agreement with Eq.~(\ref{eq:K_dimension}). Finally, we obtain
\begin{equation}
[K \phi \dot {\phi} ] = \left( \mathrm{J\,mol^{-2}} \right) \left( \mathrm{mol} \right) \left( \mathrm{mol\,s^{-1}} \right) = \mathrm{J\,s^{-1}}.
\end{equation}
Hence, the restoring pressure--volume contribution is also
dimensionally consistent with a power term.

\subsection{Particle-Transfer Contribution and the Coefficient $R$}

We next consider the dissipative particle-transfer contribution,
\begin{equation}
\Delta\mu\,\dot {N} = R \dot {\phi}^{\,2}.
\label{eq:chemical_dissipation_dimension}
\end{equation}
Since
\begin{equation}
\dot{N} = \dot{\phi},
\end{equation}
the matter-transfer rate has the dimension
\begin{equation}
[\dot N] = [\dot {\phi}] = \mathrm{mol\,s^{-1}}.
\end{equation}
Here,  the chemical potential is an energy per amount of matter. Therefore,
\begin{equation}
[\Delta\mu] = \mathrm{J\,mol^{-1}}.
\label{eq:chemical_potential_dimension}
\end{equation}
It follows that
\begin{equation}
[\Delta\mu\,\dot N] = [\Delta\mu][\dot N] =
\left( \mathrm{J\,mol^{-1}} \right)
\left( \mathrm{mol\,s^{-1}} \right) = \mathrm{J\,s^{-1}}.
\label{eq:chemical_power_dimension}
\end{equation}
Thus, the chemical-potential contribution also has the dimension of
power. On the other hand, for the right-hand side of
Eq.~(\ref{eq:chemical_dissipation_dimension}),
\begin{equation}
[\dot{\phi}^2] = \mathrm{mol^2\,s^{-2}}.
\end{equation}
Hence,
\begin{equation}
[R]\, \mathrm{mol^2\,s^{-2}} = \mathrm{J\,s^{-1}}.
\end{equation}
Solving for $[R]$ yields
\begin{equation}
[R] = \frac{\mathrm{J\,s^{-1}}} {\mathrm{mol^2\,s^{-2}}} = \mathrm{J\,s\,mol^{-2}}.
\label{eq:R_dimension}
\end{equation}

The same result follows directly from the linear-response relation
\begin{equation}
\Delta\mu=R \dot {\phi }.
\label{eq:chemical_resistance_dimension}
\end{equation}
Indeed,
\begin{equation}
[R\dot {\phi}] = \left( \mathrm{J\,s\,mol^{-2}} \right)
\left( \mathrm{mol\,s^{-1}} \right) = \mathrm{J\,mol^{-1}}.
\end{equation}
which is precisely the dimension of the chemical-potential difference
$\Delta\mu$.

Furthermore, we obtain
\begin{equation}
[R\dot {\phi}^2] = \left( \mathrm{J\,s\,mol^{-2}} \right)
\left( \mathrm{mol^2\,s^{-2}} \right) = \mathrm{J\,s^{-1}}.
\end{equation}
Therefore, one can see that the dissipative term is dimensionally consistent with a
power contribution.


\subsection{Dimensional Consistency of the Effective Thermodynamic Equation}

We take the generalized transfer coordinate $\phi$ to represent the
transferred amount of matter. Therefore,
\begin{equation}
[\phi]=\mathrm{mol}, \qquad [\dot{\phi}]=\mathrm{mol\,s^{-1}}, \qquad [\ddot{\phi}]=\mathrm{mol\,s^{-2}}.
\label{eq:phi_dimensions}
\end{equation}
The three effective coefficients have the dimensions
\begin{equation}
[M]=\mathrm{J\,s^2\,mol^{-2}}, \quad [R]=\mathrm{J\,s\,mol^{-2}}, \quad [K]=\mathrm{J\,mol^{-2}}.
\label{eq:MRK_dimensions}
\end{equation}
Their dimensional hierarchy can therefore be expressed as
\begin{equation}
[M]:[R]:[K]
\sim
\mathrm{s^2}:\mathrm{s}:1,
\label{eq:MRK_dimensional_hierarchy}
\end{equation}
apart from the common factor $\mathrm{J\,mol^{-2}}$.

The corresponding power-level contributions are
\begin{equation}
M\dot{\phi}\,\ddot{\phi},
\qquad
R\dot{\phi}^{\,2},
\qquad
K\phi\dot{\phi}.
\end{equation}
We obtained dimensions of these terms above as
\begin{align}
[M\dot{\phi}\,\ddot{\phi}]
&=
\left( \mathrm{J\,s^2\,mol^{-2}} \right)
\left( \mathrm{mol\,s^{-1}} \right)
\left( \mathrm{mol\,s^{-2}} \right) = \mathrm{J\,s^{-1}},
\\[1ex]
[R\dot{\phi}^{\,2}]
&=
\left( \mathrm{J\,s\,mol^{-2}} \right)
\left( \mathrm{mol^2\,s^{-2}} \right) = \mathrm{J\,s^{-1}},
\\[1ex]
[K\phi\dot{\phi}]
&=
\left( \mathrm{J\,mol^{-2}} \right)
\left( \mathrm{mol} \right)
\left( \mathrm{mol\,s^{-1}} \right) = \mathrm{J\,s^{-1}}.
\end{align}

Hence, we can write
\begin{equation}
[M\dot{\phi}\,\ddot{\phi}] = [R\dot{\phi}^{\,2}] = [K\phi\dot{\phi}] = [\dot U] = \mathrm{J\,s^{-1}}.
\label{eq:power_dimensional_equality}
\end{equation}
Thus, all three terms consistently belong to the same thermodynamic
power balance.

The power contributions can now be factorized as
\begin{equation}
M\dot{\phi}\,\ddot{\phi} + R\dot{\phi}^{\,2} + K\phi\dot{\phi} = \dot{\phi}
\left( M\ddot{\phi} + R\dot{\phi} + K\phi \right).
\label{eq:power_factorization_dimension}
\end{equation}

The dimensions of the three terms inside the parentheses are given by
\begin{align}
[M\ddot{\phi}]
&=
\left( \mathrm{J\,s^2\,mol^{-2}} \right)
\left( \mathrm{mol\,s^{-2}} \right) = \mathrm{J\,mol^{-1}},
\\[1ex]
[R\dot{\phi}]
&=
\left( \mathrm{J\,s\,mol^{-2}} \right)
\left( \mathrm{mol\,s^{-1}} \right)
= \mathrm{J\,mol^{-1}},
\\[1ex]
[K\phi]
&=
\left( \mathrm{J\,mol^{-2}} \right)
\left( \mathrm{mol} \right) = \mathrm{J\,mol^{-1}}.
\end{align}
Therefore, we can write
\begin{equation}
[M\ddot{\phi}] = [R\dot{\phi}] = [K\phi] = \mathrm{J\,mol^{-1}}.
\label{eq:generalized_force_dimensions}
\end{equation}
Consequently,
\begin{equation}
\left[ M\ddot{\phi} + R\dot{\phi} + K\phi \right] = \mathrm{J\,mol^{-1}}.
\label{eq:generalized_thermodynamic_force_dimension}
\end{equation}
This is precisely the dimension of a molar chemical potential,
\begin{equation}
[\mu] = [\Delta\mu] = \mathrm{J\,mol^{-1}}.
\end{equation}
Thus, the generalized quantity
\begin{equation}
M\ddot{\phi}
+
R\dot{\phi}
+
K\phi
\end{equation}
does not have the dimension of an ordinary mechanical force.
Instead, it has the dimension of a generalized thermodynamic driving
potential conjugate to the transferred amount of matter.

Accordingly, the generalized thermodynamic equation may be written as
\begin{equation}
M \ddot{\phi} + R \dot{\phi} + K \phi = \mathcal{A}(t),
\label{eq:generalized_thermodynamic_equation_dimension}
\end{equation}
where dimensional consistency requires
\begin{equation}
[\mathcal{A}] = \mathrm{J\,mol^{-1}}.
\label{eq:affinity_dimension}
\end{equation}
Hence, $\mathcal{A}(t)$ may be interpreted as a generalized
thermodynamic affinity or driving potential.

Multiplying
Eq.~(\ref{eq:generalized_thermodynamic_equation_dimension})
by the generalized flux $\dot{\phi}$ gives
\begin{equation}
M\dot{\phi}\,\ddot{\phi} + R\dot{\phi}^{\,2} + K\phi\dot{\phi} = \mathcal{A}(t)\dot{\phi}.
\label{eq:generalized_power_equation_dimension}
\end{equation}
The right-hand side has the dimension
\begin{equation}
[\mathcal{A}\dot{\phi}] = \left( \mathrm{J\,mol^{-1}} \right)
\left( \mathrm{mol\,s^{-1}} \right) = \mathrm{J\,s^{-1}},
\end{equation}
which is exactly the dimension of power,
\begin{equation}
[\dot U] = \mathrm{J\,s^{-1}}.
\end{equation}

The complete dimensional structure can therefore be summarized at
the power level as
\begin{equation}
\boxed{
\underbrace{
M\dot{\phi}\,\ddot{\phi}
}_{\mathrm{J\,s^{-1}}}
+
\underbrace{
R\dot{\phi}^{\,2}
}_{\mathrm{J\,s^{-1}}}
+
\underbrace{
K\phi\dot{\phi}
}_{\mathrm{J\,s^{-1}}}
=
\underbrace{
\mathcal{A}\dot{\phi}
}_{\mathrm{J\,s^{-1}}}.
}
\label{eq:power_level_summary}
\end{equation}
Correspondingly, at the generalized-force level,
\begin{equation}
\boxed{
\underbrace{
M\ddot{\phi}
}_{\mathrm{J\,mol^{-1}}}
+
\underbrace{
R\dot{\phi}
}_{\mathrm{J\,mol^{-1}}}
+
\underbrace{
K\phi
}_{\mathrm{J\,mol^{-1}}}
=
\underbrace{
\mathcal{A}(t)
}_{\mathrm{J\,mol^{-1}}}.
}
\label{eq:force_level_summary}
\end{equation}

Thus, both the power-level formulation and the corresponding
generalized-force formulation are dimensionally self-consistent when
$\phi$ is identified with the transferred amount of matter.


\section{Dimensionless Form of the Mechanical and LRC Systems}  \label{dim-all-sys}

The dynamical equivalence between the mechanical oscillator and the LRC
circuit becomes particularly transparent when the corresponding equations
are written in dimensionless form.

\subsection{Mechanical system}

Consider the driven damped mechanical oscillator
\begin{equation}
m \ddot{x} + b\dot{x} + k x = F(t).
\label{eq:app-clas-mot}
\end{equation}
The natural angular frequency of the system is defined as
\begin{equation}
 \omega_{m} = \sqrt{\frac{k}{m}}.
\end{equation}
Introducing the dimensionless time and displacement variables
\begin{equation}
\tau=\omega_{m} t,
\qquad
X(\tau)=\frac{x(t)}{x_{0}},
\end{equation}
where $x_0$ is a characteristic displacement scale, we have
\begin{equation}
x(t)=x_{0} X(\tau).
\end{equation}
Using $\tau=\omega_m t$, the first and second time derivatives become
\begin{equation}
\dot{x} = x_{0} \omega_m X^{\prime},
\end{equation}
and
\begin{equation}
\ddot{x} = x_{0} \omega_{m}^{2} X^{\prime \prime},
\end{equation}
where the prime denotes differentiation with respect to the dimensionless
time $\tau$.
Substituting these expressions into the equation of motion gives
\begin{equation}
m x_{0} \omega_{m}^{2} X^{\prime \prime} + b x_{0} \omega_{m} X^{\prime} + k x_{0} X = F(t).
\end{equation}
Since
\begin{equation}
m\omega_{m}^2 = k,
\end{equation}
Eq.(\ref{eq:app-clas-mot}) can be written as
\begin{equation}
k x_{0} X^{\prime \prime} + b x_{0} \omega_{m} X^{\prime} + k x_{0} X = F(t).
\end{equation}
Dividing by $k x_{0}$ yields
\begin{equation}
X^{\prime \prime} + \frac{b\omega_m}{k}X^{\prime} + X = \frac{F(t)}{k x_{0} }.
\end{equation}
Using
\begin{equation}
\frac{b\omega_m}{k} = \frac{b}{\sqrt{mk}},
\end{equation}
we define the dimensionless damping parameter
\begin{equation}
2 \zeta_{m} = \frac{b}{\sqrt{mk}},
\end{equation}
and the dimensionless driving force
\begin{equation}
f(\tau) = \frac{F(t)}{k x_{0} }.
\end{equation}
The mechanical equation therefore reduces to
\begin{equation}
X^{\prime \prime} + 2 \zeta_{m} X^{\prime} + X = f(\tau).
\end{equation}

\subsection{LRC circuit}

Consider now the series LRC circuit governed by
\begin{equation}
L \ddot{q} + R \dot{q} + \frac{1}{C}q = V(t).
\end{equation}
The natural angular frequency of the circuit is
\begin{equation}
\omega_{e} = \frac{1}{\sqrt{LC}}.
\end{equation}
We introduce the dimensionless variables
\begin{equation}
\tau = \omega_{e} t, \qquad Q(\tau) = \frac{q(t)}{q_{0}},
\end{equation}
where $q_0$ is a characteristic charge scale. Thus,
\begin{equation}
q(t) = q_{0} Q(\tau).
\end{equation}
The corresponding derivatives are
\begin{equation}
\dot{q} = q_{0} \omega_{e} Q^{\prime},
\end{equation}
and
\begin{equation}
\ddot{q} = q_{0} \omega_{e}^{2} Q^{\prime \prime}.
\end{equation}
Substitution into the LRC equation gives
\begin{equation}
L q_{0} \omega_{e}^{2} Q^{\prime \prime} + R q_{0} \omega_{e} Q^{\prime} + \frac{q_{0} }{C}Q = V(t).
\end{equation}
Since
\begin{equation}
L\omega_{e}^{2} = \frac{1}{C},
\end{equation}
we obtain
\begin{equation}
\frac{q_0}{C}Q^{\prime \prime} + R q_{0} \omega_{e} Q^{\prime} + \frac{q_0}{C}Q = V(t).
\end{equation}
Dividing by $q_0/C$ leads to
\begin{equation}
Q^{\prime \prime} + RC\omega_{e} Q^{\prime} + Q = \frac{CV(t)}{q_{0} }.
\end{equation}
Using
\begin{equation}
RC\omega_{e} = R \sqrt{\frac{C}{L}},
\end{equation}
we define the dimensionless damping parameter
\begin{equation}
2 \zeta_{e} = R \sqrt{\frac{C}{L}},
\end{equation}
and the dimensionless driving voltage
\begin{equation}
v(\tau) = \frac{CV(t)}{q_0}.
\end{equation}
The dimensionless LRC equation therefore becomes
\begin{equation}
Q^{\prime \prime} + 2 \zeta_{e} Q^{\prime} + Q = v(\tau).
\end{equation}

\subsection{Termodynamical system}

We consider the dynamical equation written in terms of the generalized
transfer coordinate $\phi$,
\begin{equation}
M \ddot{\phi} + R \dot{\phi} + K \phi = \mathcal{A}(t).
\end{equation}
In analogy with the mechanical and electrical systems, we first introduce
the characteristic angular frequency of the thermodynamical system as
\begin{equation}
\omega_{\phi} = \sqrt{\frac{K}{M}}.
\end{equation}
The corresponding dimensionless time variable is then defined by
\begin{equation}
\tau_{\phi} = \omega_{\phi} t.
\end{equation}
We also introduce a characteristic scale $\phi_c$ for the generalized
transfer coordinate and define the dimensionless variable
\begin{equation}
\Phi(\tau_{\phi}) = \frac{\phi(t)}{\phi_c}.
\end{equation}
Thus,
\begin{equation}
\phi(t) = \phi_{c} \Phi(\tau_{\phi}).
\end{equation}
Since
\begin{equation}
\frac{d}{dt} = \omega_{\phi} \frac{d}{d\tau_{\phi} },
\end{equation}
the first and second derivatives of $\phi$ become
\begin{equation}
\dot{\phi} = \phi_{c} \omega_{\phi} \Phi^{\prime},  \qquad \ddot{\phi} = \phi_{c} \omega_{\phi}^2\Phi^{\prime \prime},
\end{equation}
where the primes denote differentiation with respect to $\tau_\phi$.
Substitution into the original equation gives
\begin{equation}
M \phi_{c} \omega_{\phi}^2 \Phi^{\prime \prime} + R \phi_{c} \omega_{\phi}\Phi^{\prime} + K\phi_{c} \Phi = \mathcal{A}(t).
\end{equation}
Using
\begin{equation}
M \omega_{\phi}^{2} = K,
\end{equation}
we obtain
\begin{equation}
K \phi_{c} \Phi^{\prime \prime} + R \phi_{c} \omega_{\phi} \Phi^{\prime} + K \phi_{c} \Phi = \mathcal{A}(t).
\end{equation}
Dividing the entire equation by $K\phi_c$ yields
\begin{equation}
\Phi^{\prime \prime} + \frac{R\omega_\phi}{K}\Phi^{\prime} + \Phi = \frac{\mathcal{A}(t)} {K\phi_{c}}.
\end{equation}
Since
\begin{equation}
\frac{R\omega_\phi}{K} = \frac{R}{\sqrt{MK}},
\end{equation}
the dimensionless damping parameter can be defined as
\begin{equation}
\zeta_{\phi} = \frac{R}{2\sqrt{MK}}.
\end{equation}
Furthermore, the dimensionless driving function is defined by
\begin{equation}
\mathcal{A}_d(\tau_{\phi}) = \frac{ \mathcal{A}\left(\tau_\phi/\omega_\phi\right) } {K\phi_{c}}.
\end{equation}
The dimensionless equation therefore takes the form
\begin{equation}
\Phi^{\prime \prime} + 2\zeta_\phi\Phi^{\prime} + \Phi = \mathcal{A}_d(\tau_{\phi} ).
\end{equation}
Thus, the thermodynamical system possesses the same reduced
second-order dynamical structure as the corresponding mechanical and
electrical systems, with the characteristic frequency and damping
parameter given by
\begin{equation}
\omega_{\phi} = \sqrt{\frac{K}{M}}, \qquad \zeta_{\phi} = \frac{R}{2\sqrt{MK}}.
\end{equation}

\subsection{Common Dimensionless Dynamics}

The dimensionless equations obtained for the mechanical, electrical, and
generalized transfer systems can be written, respectively, as
\begin{equation}
X^{\prime \prime} + 2\zeta_{m} X^{\prime} + X = f(\tau),
\label{undim_mec}
\end{equation}
\begin{equation}
Q^{\prime \prime} + 2 \zeta_{e} Q^{\prime} + Q = v(\tau),
\label{undim_elec}
\end{equation}
\begin{equation}
\Phi^{\prime \prime} + 2 \zeta_{phi} \Phi' + \Phi = \mathcal{A}_d(\tau).
\label{undim_ther}
\end{equation}
Here, $X$, $Q$, and $\Phi$ denote the corresponding dimensionless dynamical
variables, while $\zeta_m$, $\zeta_e$, and $\zeta_\phi$ characterize the
dimensionless damping of the three systems.
If the dimensionless damping parameters are identified as
\begin{equation}
\zeta_{m} = \zeta_{e} = \zeta_{\phi} \equiv \zeta,
\end{equation}
and the dimensionless driving functions are chosen to have the same
functional form,
\begin{equation}
f(\tau) = v(\tau) = \mathcal{A}_d(\tau) \equiv s(\tau),
\end{equation}
all three systems reduce to the same dimensionless equation,
\begin{equation}
Y^{\prime \prime} +2 \zeta Y^{\prime} + Y = s(\tau).
\label{undim_com}
\end{equation}
The variable $Y$ therefore represents the dimensionless dynamical variable
of the corresponding physical realization, namely
\begin{equation}
Y=
\begin{cases}
X, & \text{mechanical system},\\[1mm]
Q, & \text{electrical system},\\[1mm]
\Phi, & \text{generalized transfer system}.
\end{cases}
\end{equation}
This result shows that the equivalence between these systems does not arise
from an identification of their dimensional physical variables or
parameters. Rather, it emerges only after the corresponding equations have
been expressed in dimensionless form. Quantities such as displacement,
electric charge, and the generalized transfer coordinate may have entirely
different physical meanings and dimensions, while the dimensional
coefficients governing their dynamics are likewise system dependent.
Nevertheless, after the appropriate characteristic scales are introduced,
these differences are absorbed into a small number of dimensionless
parameters.

Consequently, when the dimensionless damping parameter $\zeta$ and the
dimensionless driving function $s(\tau)$ are the same, the resulting
evolution is governed by an identical mathematical equation. The
mass--spring--damper system, the series LRC circuit, and the generalized
transfer system may therefore be regarded as distinct physical
realizations of the same linear second-order dimensionless dynamical
structure.

Importantly, this equivalence is dynamical rather than physical: it does
not imply that the underlying physical quantities are identical, but only
that their reduced evolution belongs to the same mathematical class under
the stated conditions.


\nocite{*}

\bibliography{ata.bib}.  

\end{document}